\documentclass{moriond}

\def\be{\begin{equation}}
\def\ee{\end{equation}}
\def\bea{\begin{eqnarray}}
\def\eea{\end{eqnarray}}

\newcommand{\Photo}{}

\begin{document}
\vspace*{4cm}
\title{DARK MATTER SEARCH IN THE INNER GALACTIC CENTER HALO WITH H.E.S.S.}

\author{LEFRANC. V and MOULIN. E for the H.E.S.S. collaboration}
\address{DRF/Irfu, Service de Physique des Particules, CEA Saclay, F-91191 Gif-Sur-Yvette Cedex, France}

\maketitle
\abstracts{The presence of dark matter in the universe is nowadays supported by a substantial set of astronomical and cosmological observations. A large amount of dark matter is expected in the Galactic Center (GC) region. Thanks also to its proximity, it is one of the best targets to look for dark matter particle self-annihilation into very high energy gamma-rays. We perform a search for annihilating dark matter in the central 300 parsecs around the GC with the H.E.S.S. array of ground-based Cherenkov telescopes. Using the full H.E.S.S.- I dataset (2004-2014) of GC observations, new constraints are derived on the velocity-weighted annihilation cross section $\langle \sigma v \rangle$ with a 2D likelihood method using spectral and spatial morphologies of the DM signal compared to background.  These constraints are the strongest obtained so far in the TeV mass range and  improve the previous constraints  by a factor of 5. Considering an Einasto profile, the constraints reach $\langle \sigma v \rangle$ values of $6\times 10^{-26}$cm$^{3}$s$^{-1}$ for a DM particle mass of 1.5 TeV annihilation into W$^+$W$^-$ pairs.  In the $\tau^+\tau^-$ channel, the constraints probe the natural scale for thermal relic cross section for DM particles of masses between 400 GeV and 2 TeV.}

\section{Introduction}
The presence of Dark Matter (DM) in the Universe is supported by numerous of astrophysical  and cosmological measurements, taking part of about 85\% of its mass content~\cite{Adam:2015rua}. Many well-motivated elementary particle candidates arising in extensions of the Standard Model of particle physics are proposed. One of the most promising class of models is the  Weakly Interactive Massive Particles (WIMPs). They are stable particle with masses and couplings at the electroweak scale produced in a standard thermal history of the Universe has the relic density that corresponds to that of DM, $\langle \sigma v \rangle = 3\times 10^{-26}$cm$^{3}$s$^{-1}$. The 
self-annihilation of DM particles is expected to create Standard Model particles, including gamma-rays that can be detected by ground-based Cherenkov telescopes. Overdense regions of the universe are the best place to look for a DM signal because the annihilation rate is proportional to the square of the DM density along the line of sight. The Galactic Center (GC) is one of the favorite target for indirect detection due to its proximity and the high expected DM density. Currently, the best constraints on the velocity weighted cross section above 800 GeV are obtained by H.E.S.S. and reach $\rm{3\times10^{25} \ cm^{3}s^{-1}}$ at 1 TeV for 112 hours of observation toward the GC~\cite{Abramowski:2011hc}. In this study we present an update this analysis using 10 years of data and a two dimensional likelihood to make use of the spectral and spatial characteristics of the DM compared to background

\section{Dark Matter search with H.E.S.S. in the Galactic halo}
\subsection{Galactic Center observations}
The H.E.S.S. array, located in the Khomas Highland of Namibia, is made of 4 telescopes of 12 m in diameter with a total field of view of $5^{\circ}$, and a larger one of 28 m diameter located at the center of the array aiming at lowering the energy threshold down to a few tens of GeV. Only the observation using the 4 12-m telescopes will be use in this analysis. We used for this analysis 254 hours of observations  of the GC taken between 2004 and 2014. All data pass the standard H.E.S.S. quality criteria~\cite{Aharonian:2006pe} and lead to an average zenith angle off $18^{\circ}$. The data analysis is done in a circular region of interest (RoI) of 1$^{\circ}$ radius centered on the GC excluding the Galactic latitudes $\rm{|b|<0.3^{\circ}}$ to avoid any contamination by standard astrophysical sources~\cite{Aharonian:2004wa,Aharonian:2005br,Aharonian:2006pe,Aharonian:2006au,Aharonian:2009zk}. We then subdivise the ON region into seven spatial bins corresponding to sub RoIs defined as annuli of 0.1$^{\circ}$ width centered on the GC with inner radii from 0.3$^{\circ}$ to 0.9$^{\circ}$.

\subsection{Dark matter annihilation flux}
The differential  $\gamma$-ray flux,  produced by the annihilation of self-conjugated DM particles of mass $\rm{m_{DM}}$, in a solid angle in the sky $\rm{d \Omega} = 2\pi sin\theta \rm d\theta$, is given by:
\begin{equation}
\label{promptflux}
\rm{\frac{d \Phi_\gamma^{\rm P}}{d \Omega d E_{\gamma}} = \frac{1}{8\pi m_{DM}^2}  \sum_f \langle \sigma v \rangle_f \frac{d N^f_\gamma}{d E_{\gamma}}(E_{\gamma})\times J(\theta) \ , \qquad J(\theta)=\int_{\rm l.o.s.}\rm d s \ \rho^2(r(s,\theta))} \,
\end{equation}
where  $\langle \sigma v \rangle_f$ and $d  N^f_\gamma / d E_{\gamma}$  are the thermally-averaged velocity-weighted annihilation cross section and the energy spectrum of photons in a given the channel for a finale state $f$, respectively. The coordinate $r$ writes $r(s, \theta)=(r_\odot^2+s^2-2 r_\odot s \cos\theta)^{1/2}$, where $s$ is the distance along the line of sight. $\theta$ is the aperture between the direction of observation and the GC plane, and $r_\odot=8.5$ kpc is the Sun location with respect to the GC. 
The function $J(\theta)$, or {\it J-factor}, integrates  the square of the DM density $\rho$ along the  line of sight. For the computation of the {\it J-factor}, we will assume an Einasto and an NFW profile parametrized by:
\begin{equation}
\label{profile}
\rho(r) = \rho_s  \exp \left[-\frac{2}{\alpha_s}\left(\Big(\frac{r}{r_s}\Big)^{\alpha_s }-1\right)\right] \ ,  \, \, \, \, \rho_{\rm NFW}(r) = \rho_{\rm s}\left(\frac{r}{r_{\rm s}}\Big(1+\frac{r}{r_{\rm s}}\Big)^2\right)^{-1}  \ , 
\end{equation}
The Einasto and NFW profile parameters, ($\rho_{\rm s}, \alpha_{\rm s}, r_{\rm s}$) and ($\rho_{\rm s}, r_{\rm s}$), are extracted from Ref.~\cite{Abramowski:2011hc} assuming a local DM density of $\rho_{\odot} = 0.39\ \rm GeV cm^{-3}$. The J-factors computed in the RoI, gives  $J_{\rm E} =   4.92 \times 10^{21} \ {\rm GeV^2 cm^{-5}}$ and  $J_{\rm NFW} = 2.67 \times 10^{21} \ {\rm GeV^2 cm^{-5}}$ for the Einasto and NFW profiles, respectively. An alternative parametrization of the Einasto profile~\cite{Cirelli:2010xx} leads to $J_{\rm E_2} =   1.51 \times 10^{21} \ {\rm GeV^2 cm^{-5}}$.

\subsection{Analysis methodology}
The analysis is based on a two-dimensional likelihood ratio test which uses the spatial and spectral characteristics of the DM signal versus background. For a given DM mass $m_{DM}$, the total likelihood is obtained by the product over the spatial bins $i$ and the energy bins $j$ of the individual Poisson likelihoods. It writes
$\mathcal{L}  (m_{\rm DM}, \langle  \sigma v \rangle) = \prod_{\rm i,j} \mathcal{L}_{\rm ij}$,  
with the individual likelihood is then given by~\cite{2011EPJC711554C}:
\begin{equation}
\mathcal{L}_{\rm ij}({ N_{\rm S}}, { N_{\rm B}} | { N_{\rm ON}}, { N_{\rm OFF}}, { \alpha}) = \\
\frac{\left(N_{\rm S, ij}+ N_{\rm B, ij}\right)^{N_{{\rm ON}, {\rm ij}}}}{N_{{\rm ON}, {\rm ij}}!}e^{-(N_{\rm S,ij}+ N_{\rm B, ij})} \ .
\end{equation}
$N_{\rm S, ij} + N_{\rm B,ij}$ is the expected total number of events in the spatial bin $i$ and spectral bin $j$  in the ON regions. 
$N_{\rm S, ij}$ is the expected number of signal events and   $N_{\rm B, ij}$  the number of background events expected in the spatial bin $i$ and spectral bin $j$.
$N_{\rm ON, ij}$ and  $N_{\rm OFF, ij}$ are the number of observed events in the ON and OFF regions, respectively. $N_{\rm B, ij}$ is extracted from the OFF regions and given by $N_{\rm B, ij} = \alpha_i N_{\rm OFF, ij}$. The parameter $\alpha_i=\Delta\Omega_i/\Delta\Omega_{\rm OFF}$ refers to the ratio between the angular size of the ON region $i$ and the OFF region. In our case, this ratio is equal to one since each OFF region is taken symmetrically to the ON region for each run pointing position. Consequently they have the same angular size and exposure. 
Constraints on $\langle{ \sigma v}\rangle$ are obtained from the likelihood ratio test statistic given by
$ {\rm TS}=-2 \ln(\mathcal{L}(m_{\rm DM},\langle { \sigma v} \rangle)/\mathcal{L}_{\rm max}(m_{\rm DM},\langle { \sigma v} \rangle))$, which, in the high statistics limit, follows a $\chi^2$ distribution with one degree of freedom~\cite{Rolke:2004mj}. Values of $\langle { \sigma v}\rangle$ for which TS is higher than 2.71 are excluded at 95\% confidence level (C.L.).

\section{Results}
\begin{figure}[h]
\includegraphics[width=0.45\textwidth]{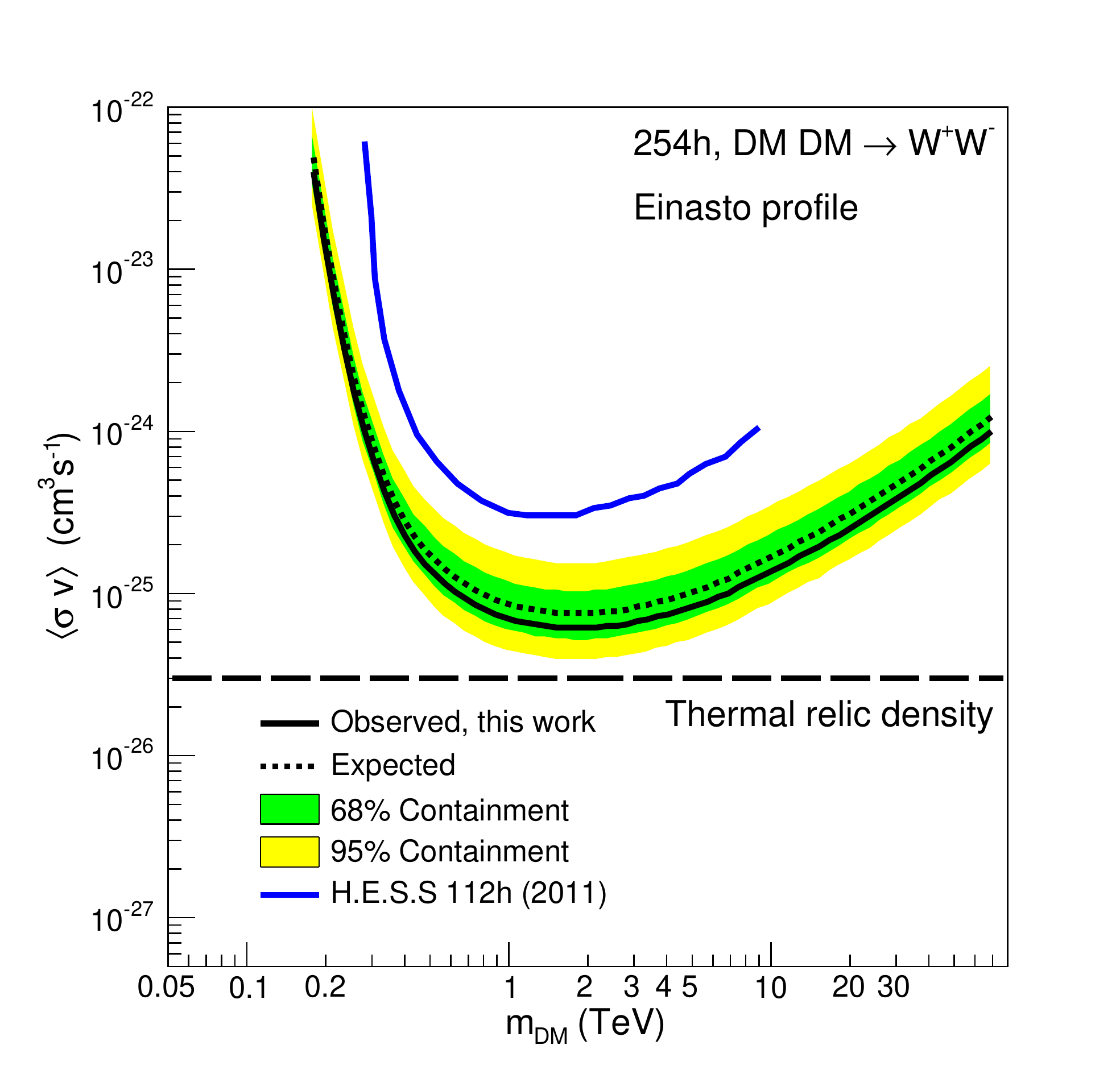}
\includegraphics[width=0.45\textwidth]{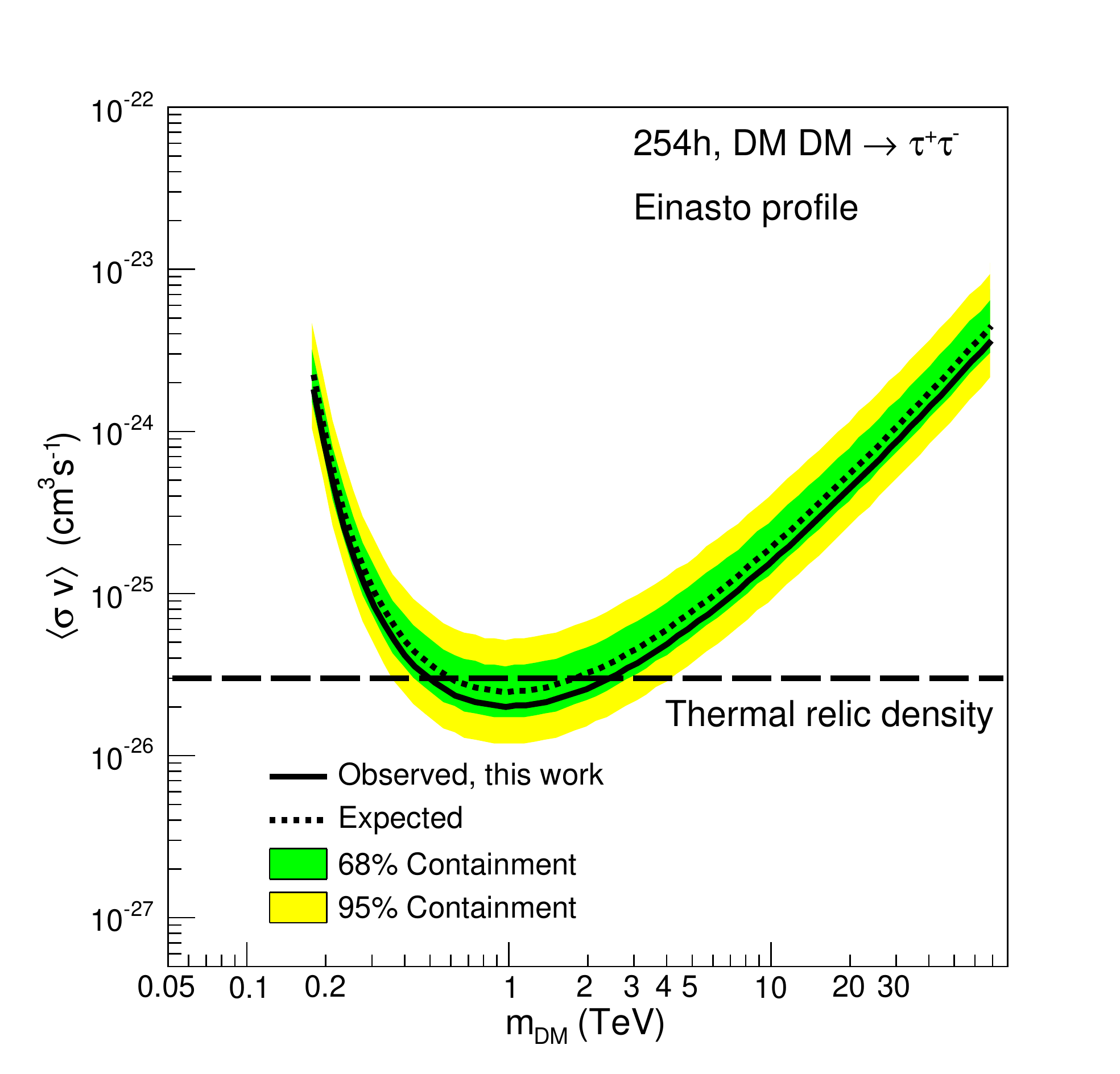}
\caption{Constraints on the velocity-weighted annihilation cross section $\langle \sigma v \rangle$ 
for the  W$^+$W$^-$ (left panel) and  $\tau^+\tau^-$ (right panel) channels derived from 10 years of observations of the inner 300 pc of the GC region with H.E.S.S. as a function of the DM mass m$_{\rm DM}$.  The observed limit is shown as black solid line.The mean expected limit (black dotted line) together with the 68\% (green band) and 95\% (yellow band) C. L. containment bands are shown. The blue solid line corresponds to the limits derived in a previous analysis of 4 years (112 h of live time) of GC observations by H.E.S.S.. 
The horizontal black long-dashed line corresponds to the thermal relic velocity-weighted annihilation cross section.}
\label{fig:results_errorJ}
\end{figure}
\begin{figure}[h]
\centering
\includegraphics[width=0.45\textwidth]{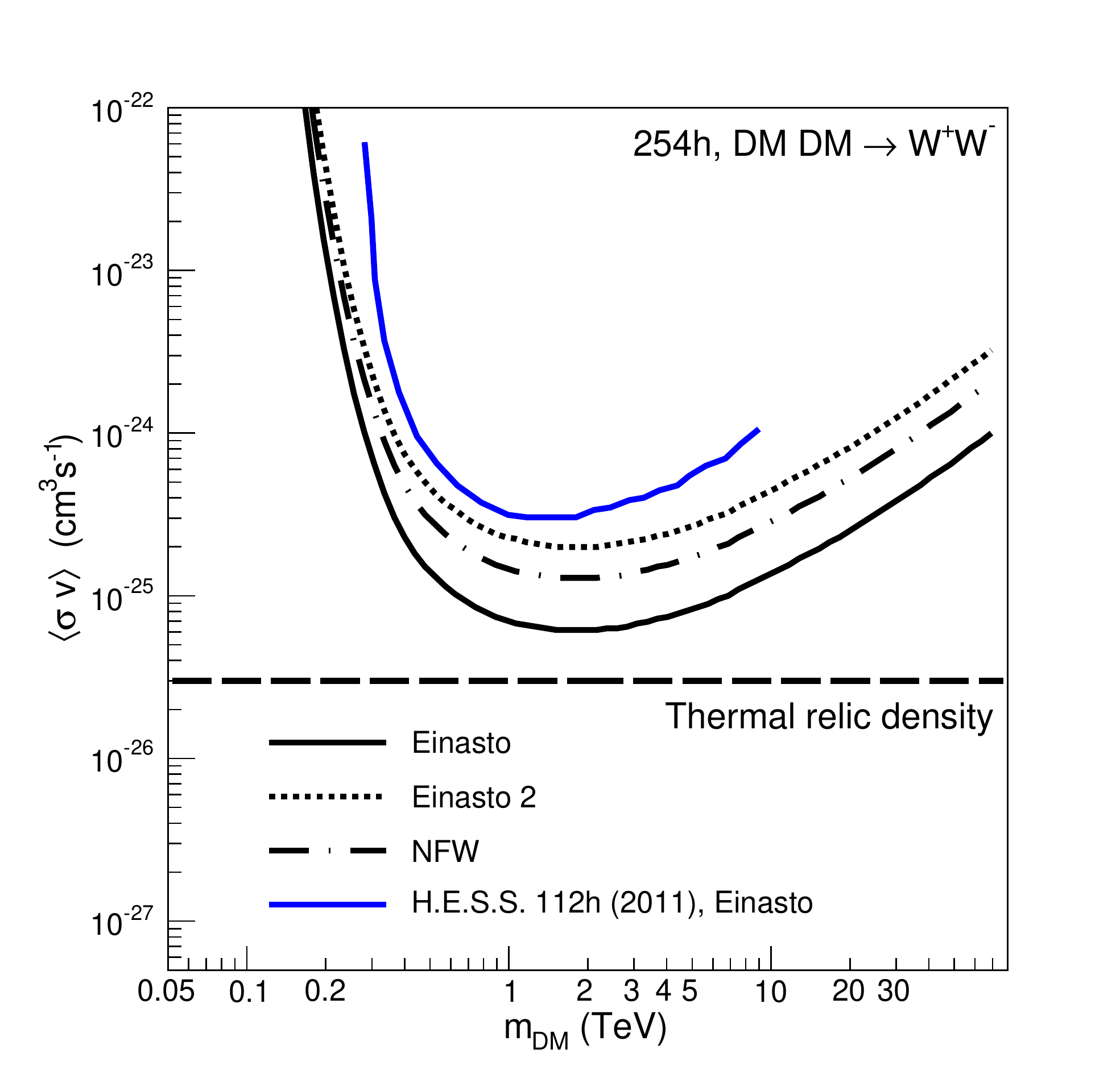}
\includegraphics[width=0.45\textwidth]{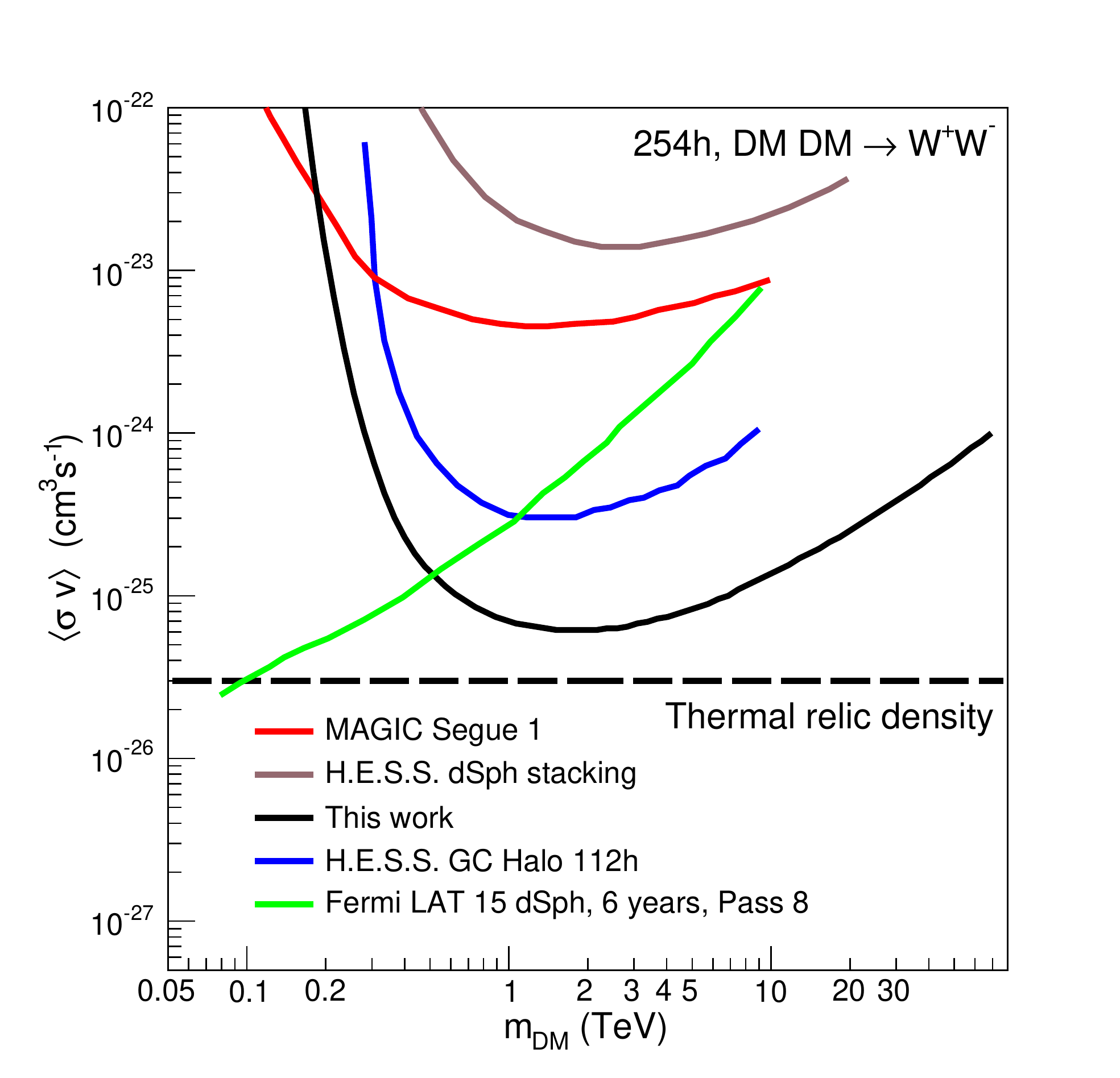}
\caption{Left: Impact of the DM density distribution on the constraints on the velocity-weighted annihilation cross section $\langle \sigma v \rangle$. The constraints expressed in terms of 95\% C. L. upper limits are  shown 
as a function of the DM mass m$_{\rm DM}$ in the W$^+$W$^-$ channels
for the Einasto profile (solid black line), another parametrization of the Einasto profile (dotted black line), and the NFW profile (long dashed-dotted black line), respectively.  
Right: Comparison of constraints on the W$^+$W$^-$ channels with the previous published H.E.S.S. limits from 112 hours of observations of the GC (blue line), the limits from the observations of 15 dwarf galaxy satellites of the Milky Way by the Fermi satellite (green line), the limits from 157 hours of observations of the dwarf galaxy Segue 1 (red line), and the combined analysis of observations of 4 dwarf galaxies by H.E.S.S. (brown line).}
\label{fig:SummaryPlot}
\end{figure}
No $\gamma$-ray excess is found in any of the RoIs. We compute the upper limits on $\langle \sigma v \rangle$ with 95 \% C. L. for WIMPs with masses between 160 GeV and 70 TeV annihilating into quark ($b\bar{b}$, $t\bar{t}$), gauge bosons (W$^+$W$^-$) and leptons ($\mu^+\mu^-$, $\tau^+\tau^-$) channels. The $\gamma$-ray spectrum coming from DM annihilation is computed using the tools available in Ref~\cite{Cirelli:2010xx}.
The left panel of Fig.~\ref{fig:results_errorJ} shows the observed 95\% C. L. upper limits for the W$^+$W$^-$ channel and the Einasto profile in black solid line. The expectations are obtained from 1000 Poisson realizations of the background measured in blank-field observations at high Galactic
latitudes. The mean expected limit (black dotted line) together with the 68\% (green band) and 95\% (yellow band) C. L. containment bands are shown. We obtain a factor 5 improvement in total compare to the previous H.E.S.S. results~\cite{Abramowski:2011hc} (blue solid line). In the right panel the same results are shown for the $\tau^+\tau^-$ channel. For the first time a ground based telescope array is able to probe the thermal relic density.
In the left panel of Fig.~\ref{fig:SummaryPlot}, the impact of the assumption on the DM distribution on the upper limit is shown for the three previous defined profile (see Eq.~(\ref{profile})).
Finally, the right panel of Fig.~\ref{fig:SummaryPlot} shows the comparaison with the more recents constraints obtain in indirect detection. We represent for the ground based telescope: MAGIC towards the dwarf galaxy Segue 1~\cite{Aleksic:2013xea}, the combined observation of 4 dwarfs galaxies by H.E.S.S.~\cite{Abramowski:2014tra} and the combined observations of 15  dwarfs galaxies by the Fermi satellite~\cite{Ackermann:2015zua}.

\section{Summary}
We performed a search for a signal of self-annihilating dark matter particles in the inner Galactic halo using the full dataset of the first phase of H.E.S.S. available. With 10 years of observations of the GC we obtained 254 hours of data. In this dataset, no significant gamma-ray excess is found in the considered RoIs, thus  we derived 95\% C.L. upper limits on $\langle \sigma v \rangle$ for DM annihilations into several channels. Together with higher
statistics and a two dimensional likelihood analysis method we improve the limits by a factor 5 compare to previous H.E.S.S. results reaching  $\rm{6 \times 10^{-26}}$ cm$^3$s$^{-1}$ for a 2 TeV DM mass in the W$^+$W$^-$ channel. The strongest limits are obtained in the $\tau^+\tau^-$ channel at $\rm 2\times10^{-26}\ cm^3s^{-1}$ for a DM particle mass of 1 TeV. For the first time, observations with a ground-based array of imaging atmospheric Cherenkov telescopes are able to probe the thermal relic annihilation cross section in the TeV DM mass range.


\section*{References}

\bibliography{bibl}

\end{document}